\documentclass[%
 reprint,
 amsmath,amssymb,
 aps,
prx,
]{revtex4-2}

\usepackage{graphicx}% Include figure files
\usepackage{dcolumn}% Align table columns on decimal point
\usepackage{amsthm}
\usepackage{braket}
\usepackage{bm}% bold math
\usepackage{float}
\usepackage[hidelinks]{hyperref}
\hypersetup{
  colorlinks   = true,
  urlcolor     = blue,
  linkcolor    = blue,
  citecolor   = blue
}

\newcommand{\rot}[1]{\mathsf{Rot}(#1)}
\newcommand{\reg}[1]{\mathsf{Reg}(#1)}
\newcommand{\dep}[2]{\text{DEP}_{#1}(#2)}
\newcommand{\depone}[1]{\dep{1}{#1}}
\newcommand{\deptwo}[1]{\dep{2}{#1}}
\newcommand{\tens}[1]{^{\otimes #1}}
\newcommand{\pidle}{p_{\text{idle}}}
\newcommand{\pinit}{p_{\text{init}}}
\newcommand{\pmeas}{p_{\text{m}}}

\begin{document}
\title{A Unitary Encoder for Surface Codes}

\author{Pei-Kai Tsai}
 \affiliation{Yale University, Department of Applied Physics, New Haven, CT 06520, USA}
 \affiliation{Yale Quantum Institute, Yale University, New Haven, CT 06511, USA}
 \author{Shruti Puri}
 \affiliation{Yale University, Department of Applied Physics, New Haven, CT 06520, USA}
 \affiliation{Yale Quantum Institute, Yale University, New Haven, CT 06511, USA}

\begin{abstract}
The surface code is a promising candidate for fault-tolerant quantum computation and has been implemented in many quantum hardware platforms. In this work, we propose a new non-local unitary circuit to encode a surface code state based on a code conversion between rotated and regular surface codes, which halves the gate count of the fastest encoder known previously. While the unitary encoders can be used to increase the code distance, the fault-distance remains fixed. Nonetheless, they can be used for space-time efficient realization of eigenstates of the surface code operators that can't be easily accessed transversally such as the Pauli $Y$-eignestate and Clifford eigenstates. It may be expected that error propagation in the non-local circuit will make decoding more challenging compared to local unitary encoding circuits. However, we find this not to be the case and that conventional matching decoders can be effectively used. Furthermore, we perform numerical simulations to benchmark the performance of our encoder against a previous local unitary encoder and the conventional stabilizer-measurement based encoder for preparing the Pauli $Y$-eigenstate and find that our encoder can outperform these in experimentally relevant noise regimes. Therefore, our encoder provides practical advantage in platforms where non-local interactions are available such as neutral atoms and trapped ions. 
\end{abstract}

\maketitle

\section{Introduction}
Quantum error correction (QEC) codes are based on encoding quantum information robustly in entangled states of multiple physical qubits and are essential for fault-tolerant quantum information. The quantum surface code is a promising QEC code where physical qubits are located on a two-dimensional square lattice with local stabilizer generators~\cite{dennis_topological_2002, kitaev_fault-tolerant_2003, fowler_surface_2012}. Due to the geometrically local structure and a high error correcting noise threshold, surface codes have been realized in diverse platforms such as superconducting qubits~\cite{google_quantum_ai_suppressing_2023}, Rydberg atomic arrays~\cite{bluvstein_logical_2024}, and trapped ions~\cite{berthusen_experiments_2024}. 

To perform any quantum computation, the first step is the preparation of logical states in QEC codes. Any target state is an eigenstate of a logical operator. In the standard approach, this target state is prepared by first initializing qubits in a product state such that it is also an eigenstate of the logical operator with the same eigenvalue as the target state. Subsequently, all the stabilizer generators are measured several times to entangle the system. Since the stabilizers commute with the logical operators, the final entangled state is also the target eigenstate. With this approach, Pauli eigenstates such as $\ket{0_L}$ or $\ket{+_L}$ can be fault-tolerantly prepared so that the logical error probability is exponentially suppressed by increasing the code size~\cite{gottesman_stabilizer_1997}. On the other hand, preparation of non-Clifford states, such as $T$-magic states $\ket T = \frac{1}{\sqrt{2}}(\ket0 + e^{i\pi/4}\ket 1)$, is not fault-tolerant, but nonetheless these faulty magic states are necessary for universal fault-tolerant quantum computation~\cite{bravyi_universal_2005}.

An alternative approach to logical state preparation of a QEC code is by a unitary encoding circuit which maps an initial product state of $k$ qubits carrying logical information along with $n-k$ ancillary qubits in known states to an $n$-qubit code state encoding the same $k$-qubit information~\cite{gottesman_stabilizer_1997}. For a surface code consisting of $n$ physical qubits, the depths of local unitary encoding circuits have a lower bound $\Omega(\sqrt n)$ due to the finite spreading speed of entanglement~\cite{bravyi_lieb-robinson_2006}, which is achieved in \cite{higgott_optimal_2021, chen_quantum_2024}. Without locality constraints on entangling gates, the unitary encoding circuits based on multi-scale entanglement renormalization ansatz (MERA)~\cite{aguado_entanglement_2008, vidal_class_2008} or graph-state decomposition~\cite{liao_graph-state_2021} can be exponentially shallower, achieving the lower bound $\Omega(\log n)$~\cite{aharonov_quantum_2018}. 

In this work, we propose a new non-local unitary encoding circuit for a $n$-qubit surface code that achieves the asymptotic circuit depth lower bound $\Omega(\log n)$~\cite{aharonov_quantum_2018} and reduces the overhead almost by a factor of two compared to previous approaches. Our protocol is motivated by a code conversion circuit between two variants (rotated and regular, as described in Sec.~\ref{sec: surface codes}) of surface codes~\cite{mcewen_relaxing_2023}. In \cite{mcewen_relaxing_2023}, the authors show that the first two steps of a depth-four stabilizer-measurement circuit can serve as a code conversion from rotated to regular codes. By applying another similar depth-two circuit on the regular code along with fresh qubits, we realize a code conversion from a regular surface code to a rotated surface code of doubled size. By combining these two circuits, we obtain a depth-four circuit that doubles the code size, surpassing the state-of-the-art MERA-based circuit which takes seven steps. 

Our unitary encoder is especially suited for Rydberg-atom qubits which have programmable long-range connectivity, as demonstrated in recent large-scale experiments, but have slower and noisier qubit measurements. The unitary encoding is a possible alternative for logical state preparation, bypassing the need for the standard stabilizer measurements. However, in the presence of noise, unitary encoders are in general less fault-tolerant, as a physical error on the qubits initially carrying the logical information can affect the final logical state, leading to a logical failure due to a single physical error. Consequently, unitary encoders may be more suitable for the preparation of non-Clifford states due to the similar error vulnerability in the standard method.

This paper is organized as follows. Section~\ref{sec: surface codes} defines the two variants of surface codes used in this work. Section~\ref{sec: unitary encoders} introduces various unitary encoding schemes and the limitations on circuit depth. Section~\ref{sec: new encoder} presents our new non-local encoder, along with numerical results comparing performance across different methods in Section~\ref{sec: results}. Finally, Section~\ref{sec: conclusion} concludes with a discussion of potential applications and future directions.

\section{Surface codes} \label{sec: surface codes}
\begin{figure}
    \centering
    \includegraphics[width=0.95\linewidth]{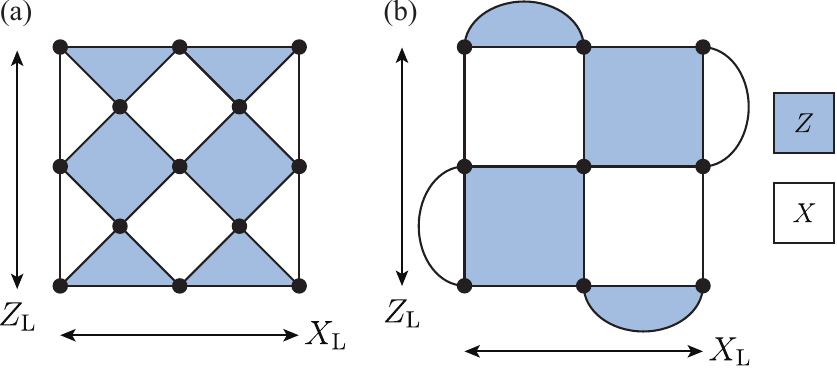}
    \caption{(a) Layout of a regular surface code. (b) Layout of a rotated surface code. Both codes have distance $d=3$. Data qubits are represented by filled circles. White (blank) and blue (filled) faces represent $X$ and $Z$ stabilizers, respectively. }
    \label{fig: surface code layouts}
\end{figure}

We focus on two variants of planar surface codes, the regular surface code and the rotated code. They are both defined on square lattices with checkerboard coloring but differ in their boundaries. In either code, qubits are placed on vertices of the lattice, and each blank (filled) face corresponds to a $X$- ($Z$-) type stabilizer generator supported on its neighboring qubits. Fig.~\ref{fig: surface code layouts}(a) illustrates a regular surface code of distance $d=3$, whose boundary creates three-body stabilizers along it. In contrast, Fig.~\ref{fig: surface code layouts}(b) shows a rotated surface code of the same distance with boundary consisting of two-body stabilizers. 

Lowest-weight X-type and Z-type logical operators $X_\mathrm{L}$ and $Z_\mathrm{L}$ are defined, respectively, as strings of Pauli $X$ and $Z$ operators along the horizontal and vertical boundaries of the code. Both codes encode a single logical qubit per code block but with different encoding rates. A regular code of distance $d$ consists of $d^2+(d-1)^2$ data qubits, while a rotated code of the same distance requires only $d^2$ data qubits. In this work, we refer to a regular surface code of size $d$ as $\reg d$ and a rotated surface code as $\rot d$. 
Note that although the code distance defined as the minimum weight of non-trivial logical operators characterizes the error tolerance of the code, in our context, a surface code of distance $d$ may not exhibit the full error-correcting capability typically associated with a distance-$d$ code. To avoid confusion, we use the notion of ``fault-distance'' to quantify the error-correcting capability throughout code operations as will be defined in Sec.~\ref{subsec: fault distance}.

\section{Unitary encoders and limitations} \label{sec: unitary encoders}
For a $k$-qubit state $\ket{\psi}$ and a stabilizer code that encodes $k$ logical qubits in $n$ physical qubits with stabilizer group $\mathcal S$, its unitary encoding circuit $U$ maps $\ket{\psi}\otimes \ket 0 \tens{n-k} $ to the $n$-qubit code state $\ket{\psi}_\mathrm{L}$~\cite{gottesman_stabilizer_1997}. Specifically, the set $\{UZ_iU^\dag\}_{i=k+1}^{n}$ generates the stabilizer group $\mathcal{S}$, and a Pauli operator $P_i$ on physical qubit $i$ ($1\leq i\leq k$, $P\in\{X, Y, Z\}$) is transformed to the corresponding logical Pauli operator $P_{i,\mathrm{L}} = UP_iU^\dag$ up to stabilizers. In this work, we study the circuit depth required to prepare code states of planar surface codes of distance $d$ with $k=1$. 

Like previous works, we start from a constant-size initial code and construct unitary circuits that grow it to distance $d$~\cite{aguado_entanglement_2008,vidal_class_2008,higgott_optimal_2021,higgott_optimal_2021, chen_quantum_2024}. Since the initial code is of constant size, it does not affect the asymptotic circuit depth and only requires a constant number of gates or measurements to prepare. The initial small code can be prepared unitarily with a different constant-depth circuit or stabilizer measurements. The latter enables post-selection on the small block which can reduce the logical error rate in the early stage of preparation. In this section, we introduce two families of encoding circuits and discuss their fault tolerance, including the advantage of initial post-selection.

\subsection{Local unitary encoders} \label{subsec: local encoder}
Due to the finite propagation speed of quantum entanglement, encoding a state into a surface code of size $d$ with local unitary circuits requires $\Omega(d)$ time steps~\cite{bravyi_lieb-robinson_2006}, which is achieved by several known constructions~\cite{higgott_optimal_2021, chen_quantum_2024}. 

\begin{figure}
    \centering
    \includegraphics[width=0.5\linewidth]{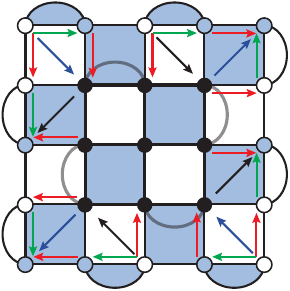}
    \caption{The gate sequence for the local encoder from a $\rot 3$ to a $\rot 5$~\cite{higgott_optimal_2021}. The original code sits in the middle. Fresh qubits are added around the boundary where the blank (filled) circles denote qubits initialized in $\ket +$ ($\ket 0$) states. Four steps of CX gates shown as arrows are applied in the order of red, green, blue, and black. The tip of the arrow is the target of the CX and the tail is the control.}
    \label{fig: local encoder}
\end{figure}

Fig.~\ref{fig: local encoder} shows a local encoding circuit~\cite{higgott_optimal_2021} that expands a surface code of size $d$ to $d+2$ in four time steps for $d=3$. The circuit begins by appending fresh qubits initialized in $\ket 0$ or $\ket +$ around the boundary of the $\rot d$ code block to form the layout of $\rot{d+2}$. These new qubits are then entangled into the code block using layers of CX gates, applied in four sequential rounds indicated by red, green, blue, and black. By repeating this process, a surface code of size $d$ can be unitarily prepared in $2d+O(1)$ time steps, saturating the Lieb-Robinson bound~\cite{bravyi_lieb-robinson_2006}. 

In the later comparison of local and non-local unitary encoders, we use the circuit in Fig.~\ref{fig: local encoder} as the implementation of local encoder.

\subsection{Non-local unitary encoders}
For general quantum circuits without locality constraints, any encoding circuit for surface codes of size $d$ requires at least $\Omega(\log d)$ time steps~\cite{aharonov_quantum_2018}. The lower bound is achieved by circuits based on multi-scale entanglement renormalization ansatz (MERA)~\cite{vidal_class_2008, aguado_entanglement_2008}, where it takes 7 time steps to double the code distance from $d$ to $2d-1$~\cite{aguado_entanglement_2008, chu_entanglement_2023}. By iteratively applying this circuit, one can prepare a surface code of size $d$ in $7\log_2 d+O(1)$ time steps. Another logarithmic-depth non-local unitary encoding circuit for a toric code based on the graph-state representations of toric code states was introduced in~\cite{liao_graph-state_2021}. This circuit has a similar depth $7\log_2d + O(1)$. 

\subsection{Fault distances of unitary encoders} \label{subsec: fault distance}
An important metric of the error tolerance of code operations is the fault distance, defined as the minimum number of physical errors in the code operation that leads to a logical error in the end. 
Although unitary encoders are able to increase the code size, in general it cannot increase the fault distance during the expanding process. This can be understood with the example: for an initial code of distance $d_\mathrm{i}$, a weight-$d_\mathrm{i}$ minimum-weight logical error is transformed to a logical error in the final code, leading to a logical error on the size-$d_\mathrm{f}$ final code due to a weight-$d_\mathrm{i}$ physical error. Thus, the fault distance of the unitary encoding process is limited by the distance of the initial code.

In non-Clifford state preparation, it is typically achieved by first prepare a small code state and grow to larger code size via stabilizer measurements~\cite{li_magic_2015,lao_magic_2022,gidney_magic_2024}. The logical error rate in this process is limited by the small code. Instead of measurements, we use unitary circuits to grow the code with the same error scaling also limited by the initial code.

\section{New encoder} \label{sec: new encoder}
\subsection{Non-local encoder based on code conversion} \label{subsec: non-local encoder}
In this section we describe our protocol to grow $\rot{d}$ to $\rot{2d-1}$. We divide our protocol into two stages.  In Stage 1, $\rot d$ is grown to $\reg d$, and in Stage 2, $\reg d$ is grown to $\rot {2d-1}$. Each stage takes two timesteps, leading to an overall four-step process transforming a $\rot{d}$ to $\rot{2d-1}$. Note that all gates in a single time step can be applied in parallel. 

\begin{figure}
    \centering
    \includegraphics[width=0.95\linewidth]{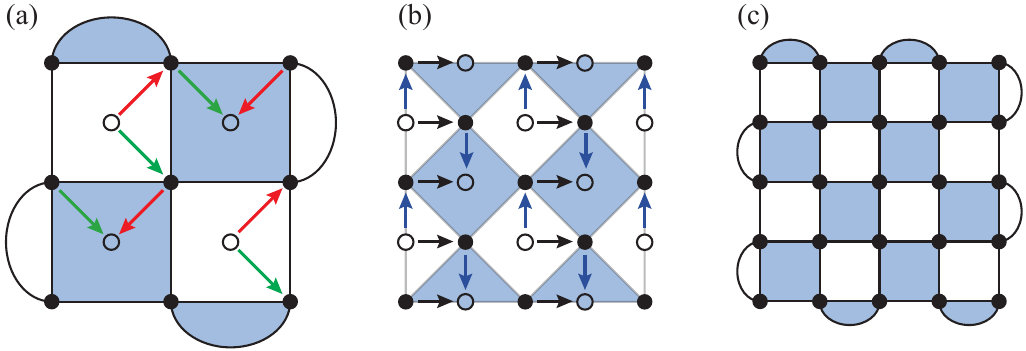}
    \caption{Gates of our non-local encoder that converts a $\rot d$ to $\reg d$ in Stage 1 and then to $\rot{2d-1}$ in Stage 2, in a total of four time steps. Four steps of CX gates shown as arrows are applied in the order of red, green, blue, and black. As before, the tip of the arrow is the target of the CX and the tail is the control. (a) The initial $\rot d$ code and gates applied in Stage 1. (b) The $\reg d$ code and gates applied in Stage 2. (c) The final $\rot{2d-1}$ code.  }
    \label{fig: my encoder gates}
\end{figure}

\textit{Stage 1}: As illustrated in Fig.~\ref{fig: my encoder gates}(a), place a qubit prepared in $\ket{+}$ on every $X$-stabilizer face and in $\ket{0}$ on every $Z$-stabilizer face in the bulk of $\rot d$. These additional qubits are shown with open circles in Fig.~\ref{fig: my encoder gates} and are referred to as ``face qubits'' associated to the corresponding faces. 
At timestep 1, apply CX gates between every face qubit and the qubit to its northeast such that the face qubit is the control (target) of the CX on the $X$- ($Z$-) stabilizer face. 
At timestep 2, apply CX gates between every $X$-stabilizer face qubit and the qubit to its southeast (face qubit as control). At the same time apply CX gates between every $Z$-stabilizer face qubits and the qubit to its northwest (face qubit as target). The gates in timestep 1 and 2 are indicated by red and green arrows respectively in Fig.~\ref{fig: my encoder gates}(a) with the tip of the arrow indicating the target and the tail as control. 

It has been shown that the state after Stage 1 is that of a $\reg d$ surface code, formed by the original $d^2$ qubits of $\rot d$ and the additional $(d-1)^2$ qubits~\cite{mcewen_relaxing_2023}. The stabilizers of resulting the $\reg{d}$ are shown as faces in Fig.~\ref{fig: my encoder gates}(b). All the qubits from Fig.~\ref{fig: my encoder gates}(a) are now data qubits in $\reg{d}$ shown as filled circles in Fig.~\ref{fig: my encoder gates}(b). Incidentally, Stage 1 is also the first half of the standard four-step stabilizer measurement circuit on a $\rot d $ code~\cite{tomita_low-distance_2014}.

\textit{Stage 2}:
Similar to Stage 1, place a face qubit prepared in $\ket{+}$ on every $X$-stabilizer face and in $\ket{0}$ on every $Z$-stabilizer face in the bulk of $\reg d$ as shown in Fig.~\ref{fig: my encoder gates}(b).
At timestep 3, apply CX gates between every face qubit and the qubit to its north such that the face qubit is the control (target) of the CX on the $X$- ($Z$-) stabilizer face. 
At timestep 4, apply CX gates between every $X$-stabilizer face qubit and the qubit to its east (face qubit as control). Also apply CX gates between every $Z$-stabilizer face qubits and the qubit to its west (face qubit as target). The gates in timestep 3 and 4 are indicated by blue and black arrows respectively in Fig.~\ref{fig: my encoder gates}(b) with the tip of the arrow indicating the target and the tail as control.

After the two stages, all $(2d-1)^2$ qubits are in the state of a $\rot{2d-1}$. By combining the above two stages we get a depth-four circuit converting a $\rot{d}$ to a $\rot{2d-1}$, allowing us to unitarily prepare $\rot{d}$ from $\rot 3$ in $4 \left(\log_2(d-1)-1\right)=4\log_2 d + O(1)$ steps for $d=2^k+1$ ($k\in\mathbb N$). Since $\rot 3$ can be prepared by a constant-depth circuit shown in Appendix~\ref{app: rot 3 prep ckt}, our proposal is a unitary encoder for $\rot d$ taking $4\log_2 d + O(1)$ timesteps. 

Note that, as suggested by Fig.~\ref{fig: my encoder gates}, although all gates in a single growth step appear local, the overall circuit becomes non-local when multiple rounds of growth are applied. As new qubits are added at the center of faces, gates from earlier stages may end up crossing over qubits added in the later stages. For example, if we further grow the $\rot 5$ code shown in Fig.~\ref{fig: my encoder gates}(c) to a  $\reg 5$ code, the new face qubit associated with the bottom-right $X$ face would sit directly in the path of the bottom-right CX gate applied in Fig.~\ref{fig: my encoder gates}(a). From the perspective of the final qubit layout, these gates in the first round are thus non-local. Consequently, in our numerical simulations, we will refer to this scheme as the ``non-local encoder'' and compare to a previously proposed local-encoder.

\subsection{Proof} \label{subsec: proof}
\begin{figure}
    \centering
    \includegraphics[width=0.95\linewidth]{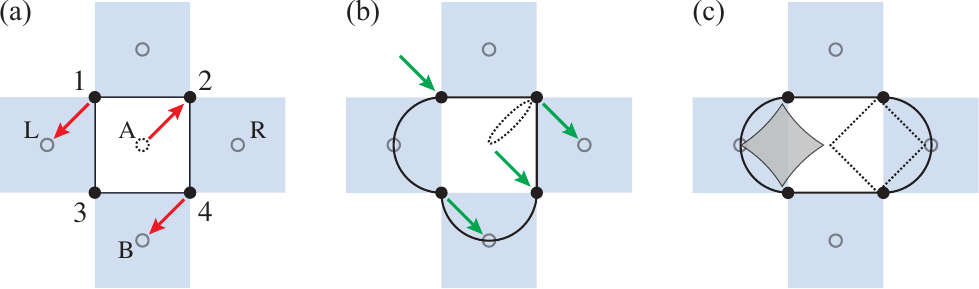}
    \caption{The evolution of rotated code stabilizers in Stage 1 of our protocol. We focus on an $X$ stabilizer in the bulk and its associated face qubit. (a) Filled circles represent data qubits supporting the $X$ stabilizer. Empty circles are the additional qubits (face qubits). Red arrows indicate the CX gates in timestep~1. (b) Support/shapes of stabilizers after gates in timestep~1, along with gates in timestep~2 shown as green arrows. (c) The final six-body (solid line) and four-body (dashed line) stabilizers. Their product is a four-body stabilizer shown as the gray square. These two square stabilizers are part of the regular code.}
    \label{fig: code conversion proof}
\end{figure}

To prove the validity of our protocol we simply need to track how the stabilizers and logical operators evolve through the circuit. In this direction, consider Fig.~\ref{fig: code conversion proof} which shows four data qubits $(1,2,3,4)$ supporting an $X$-stabilizer and its associated face qubit $\mathrm A$, and the neighboring face qubits $(\mathrm L, \mathrm R, \mathrm B)$ to its left, right, and bottom for a $\rot d$ respectively. Qubit indices in subscripts denote tensor products, e.g. $X_{123}=X_1\otimes X_2\otimes X_3$. 

We first consider a code stabilizer $X_{1234}$ and the stabilizer $X_\mathrm{A}$ of its face qubit. They are transformed to $X_{\mathrm{1234LB}}$ and $X_\mathrm{A2}$ by gates in the first timestep (red), then to $X_{\mathrm{1234LR}}$ and $X_{\mathrm{2A4R}}$ by gates in the second timestep (green). These are equivalent to $\{X_{\mathrm{2A4R}}, X_{\mathrm{1234LR}}X_{\mathrm{2A4R}}\}=\{X_{\mathrm{1L3A}}, X_{\mathrm{2A4R}}\}$, forming two neighboring $X$-stabilizers in a $\reg d$ code. For the stabilizers on the boundaries of $\rot d$, the same analysis shows that they are transformed to boundary stabilizers of $\reg d$ (see Appendix~\ref{app: stabilizer proof} for details). For $Z$ stabilizers, it can be similarly shown that a $Z$-stabilizer and its face qubit in a $\rot d$ are transformed to two neighboring $Z$-stabilizers in a $\reg d$. 

\begin{figure}
    \centering
    \includegraphics[width=0.85\linewidth]{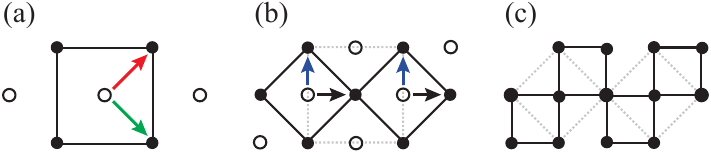}
    \caption{The evolution of a $X$ stabilizer and relevant face qubits during a $\rot d$-to-$\rot{2d-1}$ expansion. (a) A $X$ stabilizer (solid square) and its face qubits (middle circle) in a $\rot d$ code. The arrows represent gates applied in Stage 1. (b) The stabilizer and face qubit in (a) become two stabilizers of a $\reg d$ code (two tilted squares). Blank circles and arrows represent new face qubits added and gates applied to the code in Stage 2, respectively. (c) The two $X$ stabilizers and their face qubits are transformed into four smaller squares of halved face size, forming a $\rot{2d-1}$ code. }
    \label{fig: code conversion proof 2}
\end{figure}

Fig.~\ref{fig: code conversion proof 2} illustrates the growth process described above. Let $a$ denote the side length of an $X$-stabilizer in the $\rot d$ code shown in Fig.~\ref{fig: code conversion proof 2}(a). After Stage 1, each $X$ stabilizer and its associated face qubit are transformed to two smaller squares, rotated by $45^\circ$ and with side length $a/\sqrt 2$, which share a common vertex at the original face qubit position as shown in Fig.~\ref{fig: code conversion proof 2}(b). 

Observe that the bulk stabilizers and gates in Stage 2 have the same structure as those in Stage 1 after a $45^\circ$ rotation of the original lattice. The length of faces in Stage 2 is shrunk by a factor of $\sqrt{2}$ compared to that in Stage 1, but this shrinking is inconsequential for stabilizer evolution. Thus, we can simply extend the stabilizer transforms of Stage 1 to see that after Stage 2, each $X$ stabilizer in the $\reg d$ is further shrunk and rotated into squares with side length $a/2$, as shown in Fig.~\ref{fig: code conversion proof 2}(c). These smaller faces form the stabilizer group of a rotated surface code with halved face size, or equivalently, a doubled code distance. 

For the boundary stabilizers and the logical operators, a similar stabilizer-tracking argument shows that they are mapped to a $\rot{2d-1}$, as detailed in Appendix~\ref{app: stabilizer proof}. Therefore, the four-step gate sequence in Fig.~\ref{fig: my encoder gates} transforms a $\rot d$ code to a $\rot{2d-1}$ code, resulting in a unitary circuit of depth $4\log_2 d+O(1)$ to grow a $\rot 3$ to a $\rot d$. It has a $\frac47\times$ shorter circuit compared to the MERA-based non-local encoders~\cite{aguado_entanglement_2008, chu_entanglement_2023}. \qed

\section{Results} \label{sec: results}
In this section we present numerical results to benchmark and compare the performances of three different state encoding schemes: (a) our non-local encoder, (b) a local-encoder~\cite{higgott_optimal_2021}, and (c) the conventional approach based on stabilizer measurements~\cite{li_magic_2015}. All the three approaches are based on preparation of an initial $\rot{3}$ code and then expansion to a larger code of distance $d_\mathrm{f}$. For all the three approaches, we chose to prepare the $\rot{3}$ code state using the standard stabilizer measurement approach with two rounds of measurement and post-selection~\cite{li_magic_2015} (see Appendix~\ref{app: initial prep} for more details). In approach (c) the $\rot{3}$ code is grown to a distance $d_\mathrm{f}$ code by two additional rounds of stabilizer measurements \cite{li_magic_2015} (see Appendix~\ref{app: initial prep}).

To extract the logical error rate for the unitary encoders in numerical simulations, we perform one round of perfect syndrome measurements and use PyMatching~\cite{higgott_sparse_2023} to implement a minimum-weight perfect matching (MWPM) decoder on the resulting syndromes. This is similar to how logical error rates are extracted for the conventional approach. 
Although we think the main benefit of using unitary encoders is for preparing magic states, in this paper we simulate the preparation of the +1 $Y$-eigenstate $\ket{+i}$ using Stim~\cite{gidney_stim_2021}. We chose this state since it is a Pauli-eigenstate making Stim simulations easy. Moreover, like the magic state, this state is also hard to prepare fault-tolerantly on the surface code and degrades due to both logical $X$ and $Z$ errors like the magic state. Thus, our numerical simulations provide a good benchmark for how the different encoders perform. 

\subsection{Noise model and decoder}
To simulate the effects of noise, we apply one- and two-qubit depolarizing channels on qubits, where a $t$-qubit depolarizing channel of noise strength $p$, denoted by $\dep{t}{p}$, stochastically applies a Pauli error sampled from $\{I,X,Y,Z\}\tens t \setminus \{I\tens t\}$, each with probability $p/(4^t-1)$, on its support. In the simulations, we apply circuit-level noise models parametrized by four noise strengths $(p_2, \pidle, \pinit, \pmeas)$ where (i) each two-qubit gate is followed by a $\deptwo{p_2}$, (ii) each idle qubit experiences a $\depone{\pidle}$, (iii) each fresh qubit added into the system for the next round of expansion has an initialization error $\depone{\pinit}$ and (iv) each single qubit measurement is preceded by $\depone{\pmeas}$. Unless otherwise specified, we set $\pidle=\pinit=p_1$ for simplicity, in which case the noise model is characterized by three parameters $(p_1, p_2, \pmeas)$.

A nice property of the surface code is that a single $X$ or $Z$ error anticommutes with a pair of stabilizers allowing for efficient decoding with MWPM. Fortunately, we find that despite the possible long-range error propagation in our unitary encoder, a single $X$ or $Z$ error propagates to a Pauli error string that still only anti-commutes with a pair of stabilizers (See Appendix~\ref{app: decoder} for details). Thus, we can still use MWPM to effectively decode the unitarily-prepared surface code under the circuit level noise model described above.

\subsection{Perfect initial code state}
To isolate and understand the effect of errors during the unitary expansion process, we first consider a perfect $\rot{3}$ code and expand it to a final distance $d_\mathrm{f}=2^m+1$ ($m\in\mathbb N$) using the local encoder (Sec.~\ref{subsec: local encoder}) or our non-local encoder (Sec.~\ref{subsec: non-local encoder}). 

For a fixed final code distance, the circuit depth of the local encoder is much larger than that of the non-local one. So we expect non-local encoder to yield a much lower logical error rate than the local one as the distance increases because the overall circuit noise will be lower in case of the non-local encoder. However, note that, in the local encoder the bulk qubits are only affected by idle errors, while boundary qubits are affected by gate errors. This is in stark contrast with our non-local encoder where all the qubits are acted upon by gates in each growth step. Thus, we expect the local encoder to perform better than the non-local one when idle errors are negligible compared to gate errors. As a result of this dependence on $p_1/p_2$ as well as total circuit-depth, we expect to see a cross-over ratio $p_1/p_2$ when one encoder does better than the other. At lower $p_1/p_2$ the local encoder does better but as $p_1/p_2$ increases the noise due to larger circuit depth catches up making it worse than the non-local encoder. For this reason we also expect the crossover ratio to also decrease with increasing final code distance. 

This behavior is confirmed by Fig.~\ref{fig: p1, p2 ratio comparison} which compares the dependence of the logical error rates on $p_1/p_2$ at different final code distances. Consistent with our analysis, the non-local encoder outperforms the local one once single-qubit error is strong enough. However, the crossover $p_1/p_2$ when the non-local encoder yields a lower logical error rate than the local one is lower for larger final distance. Thus, our encoder outperforms the local one over a wider range of noise strength at larger final distances.

\begin{figure}
    \centering
    \includegraphics[width=0.9\linewidth]{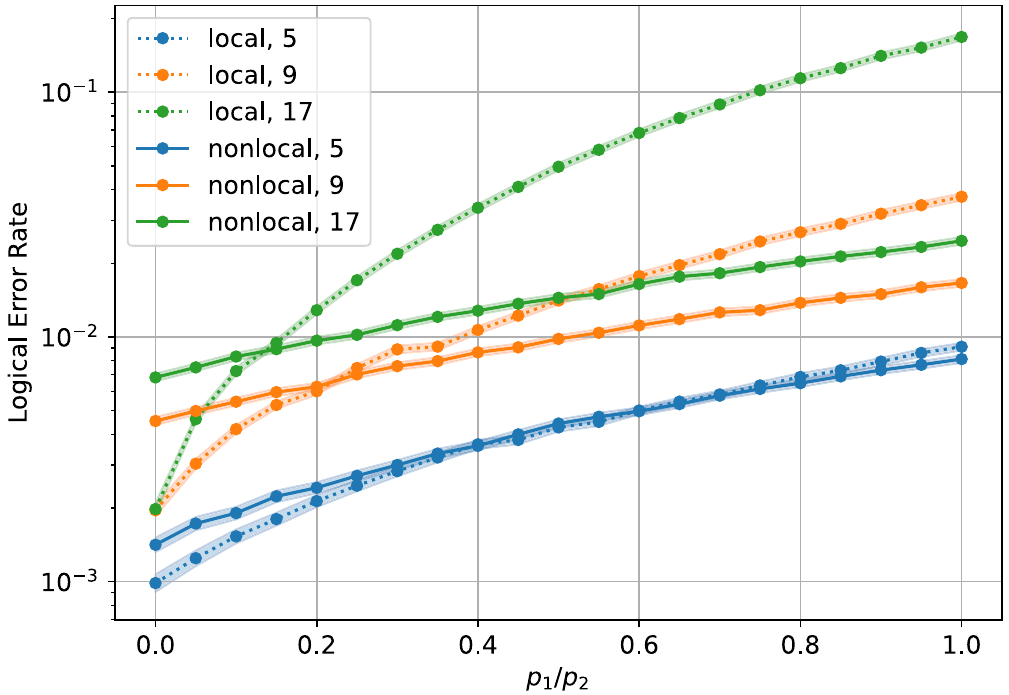}
    \caption{Comparison of logical error rates for different $p_1/p_2$ ratios under non-local and local unitary encoders. A perfect initial code block and gate error $p_2=0.5\%$ are assumed. Dashed line and solid line represent the local and non-local encoder. Blue, orange, and green lines represent the final distance $d_\mathrm{f}=5, 9, 17$, respectively. }
    \label{fig: p1, p2 ratio comparison}
\end{figure}

\subsection{Imperfect initial code state} \label{subsec: end-to-end}
\begin{figure}
    \centering
    \includegraphics[width=0.9\linewidth]{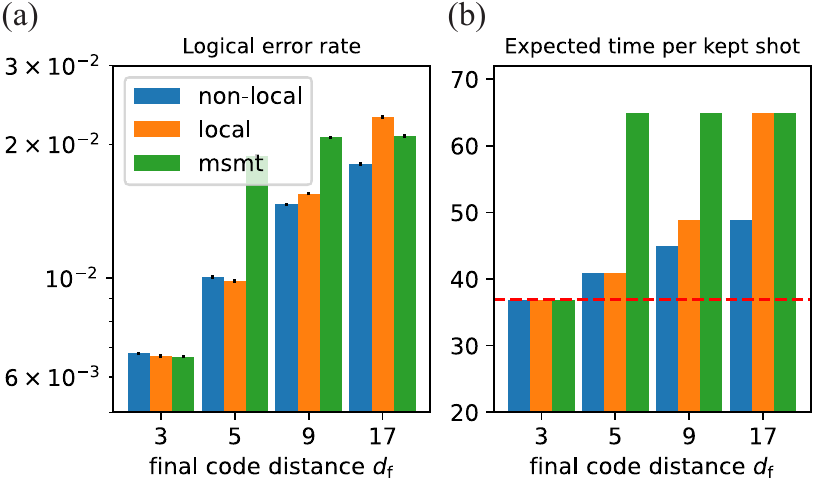}
    \caption{An end-to-end comparison of non-local unitary, local unitary and the standard encoders of the $Y$-state preparation. We assume $p_2=\pmeas=5p_1=0.5\%$ and $T_\mathrm{m}/10 = T_\mathrm{2q} = 1$ in our noise model. (a) Logical error rates of different final code distances and encoders. (b) Time overhead of different final code distance and encoders. The red dashed line represents the preparation time of the initial code block.}
    \label{fig: three encoders comparison by time}
\end{figure}

We now consider the realistic scenario where the initial $\rot{3}$ code is also prepared using noisy gates and measurements and compare the performance of the three encoding schemes in Fig.~\ref{fig: three encoders comparison by time}. We use a noise model where $p_2=\pmeas=5p_1=0.5\%$, reflecting the fact that two-qubit and measurement errors are typically stronger than single-qubit error in most hardware. At these physical error rates, the acceptance rate, $p_\mathrm{succ}$, for the post-selected preparation of the initial $\rot{3}$ code is $p_\mathrm{succ}=0.759(2)$. 

As shown in Fig.~\ref{fig: three encoders comparison by time}(a), the logical error rate for the conventional stabilizer measurement approach remains roughly constant across different final code distances $d_\mathrm{f}\geq 5$, as the overall error is dominated by the initial code and the overall preparation depth is fixed (four rounds of stabilizer measurements). On the other hand, the logical error rates of unitary encoders grow as the code distance increases because the circuit depth increases. Remarkably however, we find that for small, but yet experimentally relevant, final code distances ($d_\mathrm{f}\lesssim 9$), both unitary encoding schemes can outperform the conventional approach, with the non-local one yielding the lowest logical error rates. The non-local encoder also yields the lowest logical error rate at $d_\mathrm{f}= 17$, while the local encoder does worse than the conventional approach. Note that for $d_\mathrm{f}=3$, no expansion is needed, so the three encoding schemes have identical logical error rates. 

The most significant advantage of the non-local encoder, however, is that it can generate the states faster than the other encoders. This is demonstrated by Fig.~\ref{fig: three encoders comparison by time}(b) which compares the expected preparation time per successful attempt for the three encoders. Here we assume that a measurement is an order of magnitude longer than a two-qubit gate, i.e., $T_\mathrm{m}/10 = T_\mathrm{2q} = 1$ which is consistent with many experimental platforms~\cite{ma_universal_2022}. The initial $\rot{3}$ code preparation takes an expected time of $T_{\rot{3}}=2(4T_\mathrm{2q} + T_\mathrm{m})/p_\mathrm{succ}$. For both unitary encoders, each growth step adds $T_\mathrm{step}=4T_\mathrm{2q}$ to the total time required to grow to $d_\mathrm{f}$, while the conventional method requires $T_\mathrm{LS}=2(4T_\mathrm{2q} + T_\mathrm{m})$ for the two rounds of stabilizer measurements required to grow $\rot{3}$ to $\rot{d_\mathrm{f}}$. 

Clearly, the non-local encoder requires fewer steps than the local one to reach the target code distance and so will always have a lower expected time per kept shot. Remarkably, however, at $d_\mathrm{f}\lesssim 17$, the non-local unitary encoder is also faster than the conventional approach which is slowed down by the length measurements.

This end-to-end comparison shows that the non-local unitary encoder can outperform the local unitary encoder, as well as the conventional encoder, not only by achieving a lower logical error rate at the same final code distance, but also by requiring less preparation time. Note that the above preparation time of the initial code is calculated under the assumption of operations on a single code block and a repeat-until-success strategy, leading to the $1/p_\mathrm{succ}$ factor in the expression for $T_{\rot{3}}$. However, a space-time tradeoff can be made for the initial $\rot{3}$ code reducing the time per kept shot in every approach.

\section{Conclusion and Discussion} \label{sec: conclusion}
In this work, we propose a non-local unitary encoding circuit for surface codes based on a code conversion between regular and rotated surface codes, achieving approximately $50\%$ reduction in circuit depth compared to the previous best non-local encoders based on MERA circuits or graph-state decomposition. We furthermore benchmark the performance of our encoder with the standard stabilizer-measurement method and a local unitary encoder, and find numerically that our encoder can outperform both of them, yielding lower logical error rates and faster generation rates in an experimentally relevant noise regime.

Our protocol is best suited for platforms with non-local two-qubit interactions, such as trapped ions and neutral atoms. In the neutral atom platform, for example, one can begin with a $\rot{3}$ code on a large lattice and bring in the additional qubits on demand via movable acoustic-optical deflectors (AODs). Then the two stages of our protocol can be implemented which are essentially the first half of a stabilizer measurement cycle in the surface code. Thus, existing machinery optimized for surface code stabilizer measurements can easily be used with our scheme~\cite{bluvstein_logical_2024}. This is in direct contrast to the local unitary encoding circuit in Fig.~\ref{fig: local encoder} where the gate ordering is considerably different from the conventional stabilizer measurement procedure. Moreover, measurements can be more than an order of magnitude slower than two-qubit gates in neutral atom platforms~\cite{ma_high-fidelity_2023, bluvstein_logical_2024}. Additionally, depending on how the gates are composed, idle errors on qubits not participating in the gates can be comparable to the two-qubit gate error within an order of magnitude~\cite{bluvstein_quantum_2022, ma_high-fidelity_2023, bluvstein_logical_2024}. Thus, the end-to-end comparison is Section~\ref{subsec: end-to-end} demonstrates that our non-local protocol can outperform the stabilizer measurement and local encoding approach in neutral atom qubits. This numerical analysis motivates further calculations with detailed accounting of atom movements and noise to identify the winning scheme. 

Our protocol opens up additional promising directions for future work. Firstly, it can be useful for generation of low-error magic states via cultivation~\cite{gidney_magic_2024} or distillation~\cite{bravyi_universal_2005}. In particular, magic state cultivation starts with a noisy copy of a magic state and gradually grows its size and logical fidelity. Figure~\ref{fig: three encoders comparison by time} shows that errors introduced during growing from a $d=3$ to a $d_\mathrm{f}$ code via stabilizer measurements can be much larger than the added error rate per growing step with the non-local unitary encoder. Thus, depending on the starting and final code distances, the unitary encoders can be more effective for magic states cultivation compared to the conventional approach. We will consider the application of our encoder for magic state cultivation in a future work~\cite{msc-in-prep}. 

Furthermore, it will be useful to investigate how to adapt our proposed unitary encoding technique to prepare other topological codes, in particular the color code since there is a known local unitary conversion circuit between color and surface codes~\cite{kubica_unfolding_2015}. Finally, it will be useful to consider application of the non-local encoder to prepare ansatz states for quantum algorithms, like quantum phase estimation for ground state energy estimation. The requirement for the ansatz state is that it must have a non-exponentially-small overlap with a desired target state and therefore it may be acceptable to prepare these starting from non-fault-tolerant states~\cite{ding_even_2023, ni_low-depth_2023, wang_quantum_2023}.

\begin{acknowledgements}
This material is based upon work supported by the U.S. Department of Energy, Office of Science, National Quantum Information Science Research Centers, Co-design Center for Quantum Advantage (C\textsuperscript{2}QA) under contract number DE-SC0012704. We thank Jahan Claes, Thomas Smith, Jeff Thompson, and Bichen Zhang for useful discussions and feedback. 
\end{acknowledgements}

\appendix
\section{A unitary circuit encoding $\rot 3$ code}\label{app: rot 3 prep ckt}
\begin{figure}
\centering
\includegraphics[width=0.9\linewidth]{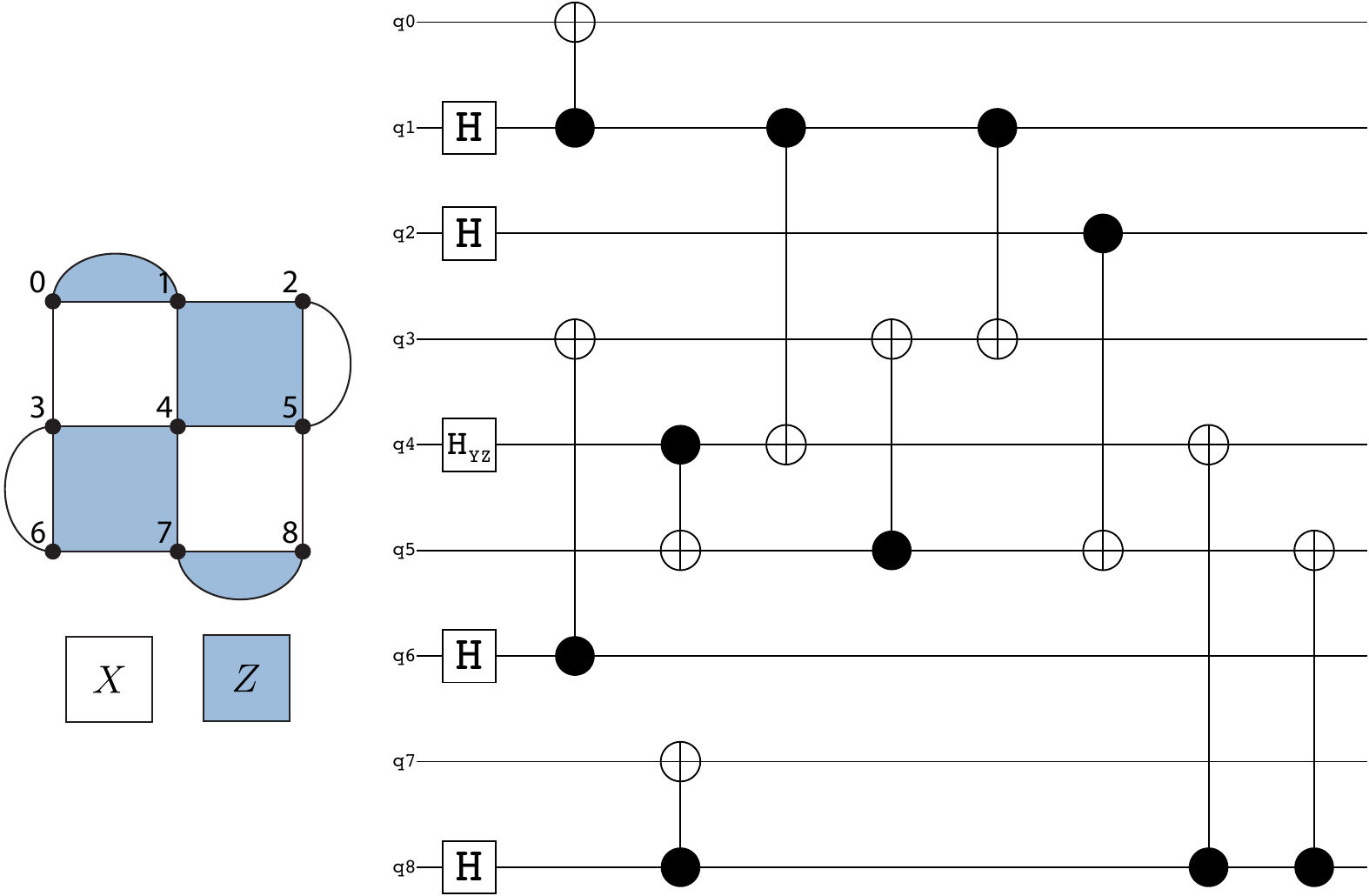}
\caption{A unitary encoding circuit for a $\rot 3$ code preparing $\ket {+i_\mathrm{L}}$ logical state. Qubit indices in the circuit are shown on the left code block. The circuit is drawn by Stim~\cite{gidney_stim_2021}.}
\label{fig: rot 3 prep ckt}
\end{figure}

Fig.~\ref{fig: rot 3 prep ckt} shows a unitary encoding circuit for the initial $\rot 3$ code. The qubit ``q4'' (qubit in the middle of the code block) carries the logical information. The $H_\mathrm{YZ}$ gate acting on the initial $\ket 0$ state produces $\ket{+i}$ state for the simulations. For magic state preparation, it should be replaced by a $TH$ gate such that it encodes $TH\ket 0 = \ket T$. 

\section{Proof of code conversion for stabilizers on the boundaries and logical operators} \label{app: stabilizer proof}

\begin{figure}
    \centering
    \includegraphics[width=0.9\linewidth]{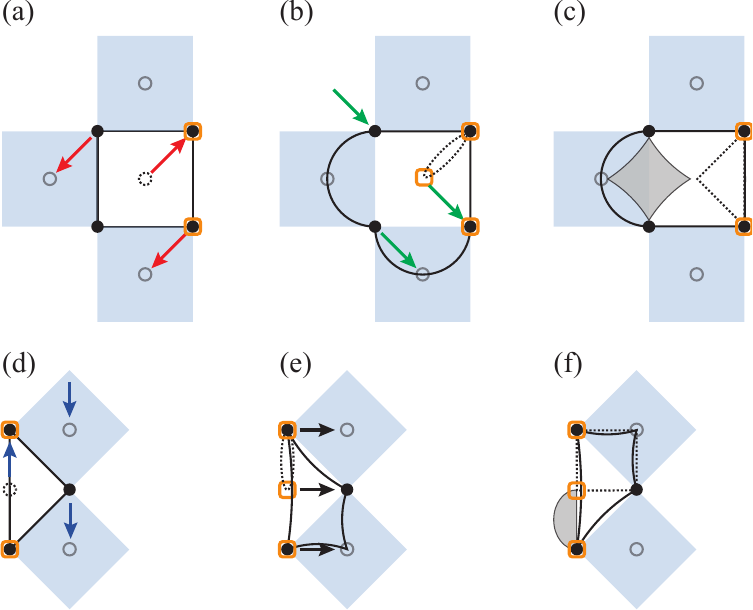}
    \caption{Proof of code conversion for stabilizers on the boundaries and logical operators}
    \label{fig: boundary proof}
\end{figure}

To complete the proof in Sec.~\ref{subsec: proof}, here we show that the stabilizers near the code boundaries are transformed correctly. For example, Fig.~\ref{fig: boundary proof}(a) highlights an $X$ face and its associated face qubit near the $X$-type right boundary. After two timesteps of gates in Stage 1 (Fig.~\ref{fig: boundary proof}(a)-(c)), the original $X$ stabilizer and its face qubits are transformed into two smaller rotated stabilizers, where the one in the bulk is a rotated square similar as before, while the other one close to the boundary is cut into a triangle. The latter forms the weight-three stabilizers along the boundaries of a regular surface code. The stabilizers close to other boundaries are similarly transformed into one bulk square and one weight-three face on the boundary of a $\reg d$ code. Note that every weight-two stabilizer on the boundaries of the $\rot d$ code is touched by a single gate that propagates the stabilizer to a face qubit, forming a weight-three boundary stabilizer of the $\reg d$ code. 

Fig.~\ref{fig: boundary proof}(d)-(f) show the evolution of an $X$ stabilizer on the left boundary of a $\reg d$ code. Instead of being transformed into two smaller square faces as in Fig.~\ref{fig: code conversion proof 2}, it becomes one square and one weight-two boundary stabilizer of the $\rot{2d-1}$ code. The stabilizers along other boundaries are transformed similarly. 

In Fig.~\ref{fig: boundary proof}(a), qubits supporting a logical operator $Z_\mathrm{L}$ along the right boundary are highlighted by orange boxes.  After applying the gates in the first stage, the support of $Z_\mathrm{L}$ remains unchanged and still represent a logical $Z_\mathrm{L}$ in the resulting $\reg d$ code. Orange boxes in Fig.~\ref{fig: boundary proof}(d)-(f) show the evolution of a logical operator $Z_\mathrm{L}$ along a boundary of a $\reg d$ code to a $\rot{2d-1}$ code. By the final stage, the operator includes additional support on face qubits from the $\reg d$ code, indicating that its weight (and thus the code distance) has increased.

\section{State preparation by noisy stabilizer measurements} \label{app: initial prep}
\begin{figure}
    \centering
    \includegraphics[width=0.45\linewidth]{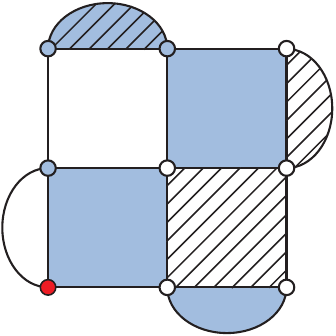}
    \caption{The initial qubit state for injecting an arbitrary state into a surface code block. The bottom-left red qubit carries the logical information to be injected. White (blue) qubits are prepared in $\ket +$ ($\ket 0$) state. In the first round of stabilizer measurements, all shaded stabilizers should read +1 and thus are capable of detecting errors in the first round.}
    \label{fig: product state pattern}
\end{figure}
The standard way of preparing the logical code state $\ket{\psi_L}$ by stabilizer measurements starts with initializing data qubits in certain product states shown in Fig.~\ref{fig: product state pattern}, where the bottom-left qubit in state $\ket\psi$ carries the logical information, and qubits above (below) the dashed diagonal are in $\ket +$ ($\ket 0$) states~\cite{li_magic_2015, lao_magic_2022}. This initial state is an eigenstate of $\alpha X_L+\beta Y_L + \gamma Z_L$ if $\ket \psi$ is an eigenstate of $\alpha X+\beta Y + \gamma Z$. By performing multiple rounds of stabilizer measurements, the system is projected into logical state $\ket{\psi_L}$ with error-detecting capability. In the first round of measurements, shaded stabilizers in Fig.~\ref{fig: product state pattern} should give definite measurement outcomes. The subsequent rounds should have outcomes consistent with previous rounds. We discard the shots that violate the above conditions. This scheme has an overall $O(p)$ logical error rate similar to pure unitary encoders because the initial state pattern leaves the physical qubit initially containing logical information completely unprotected. Note that in the quantum circuit of the first round of stabilizer measurement, we remove trivial gates such as a CX whose control qubit is initialized in $\ket 0$ to reduce the amount of gate error~\cite{gidney_magic_2024}.

In the end-to-end comparison (Sec.~\ref{subsec: end-to-end}), the initial code is prepared by two rounds of noisy stabilizer measurements followed by post-selection. The choice of two rounds is motivated by numerical simulations. A single round of measurements results in an additional $O(\pmeas)$ contribution to the logical error rate due to measurement errors. This does not change the overall $O(p)$ scaling set by the initial unprotected product state, but it increases the constant prefactor significantly. Simulations indicate that two rounds of measurement reduce the logical error rate by a factor of four compared to a single round. On the other hand, more than two measurement rounds does not further suppress the logical error rate, as it is still dominated by the initial $O(p)$ contribution. 

When the initial code is grown to the final code distance via the standard stabilizer measurements, we apply two additional rounds of stabilizer measurements of the final code for code-space projection and mitigating first-order logical errors due to measurement errors, followed by a final round of perfect stabilizer measurements for decoding.

\section{Error propagation and matchable decoding graph}\label{app: decoder} 
To justify the use of matching-based decoders, one must show that each error mechanism either triggers at most two detection events or is decomposable into such components. While we do not provide an analytical proof of this condition for our circuits, we verify it numerically. For both the local and our non-local encoder, we numerically confirm that all error mechanisms under a circuit-level noise model lead to either edges or decomposable hyperedges (as in the case of $Y$ errors) in the decoding graph. This structure supports the applicability of the minimum-weight perfect matching (MWPM) decoder~\cite{higgott_sparse_2023}.

\begin{figure}[h]
    \centering
    \includegraphics[width=0.95\linewidth]{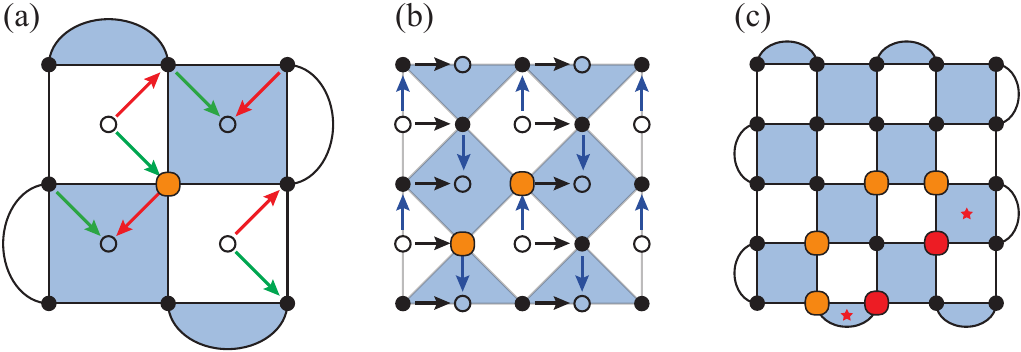}
    \caption{Error propagation in a single round of expansion with our non-local encoder.}
    \label{fig: error propagation}
\end{figure}

For example, Fig.~\ref{fig: error propagation} illustrates how a single error propagates through one round of expansion. Assume an $X$ error (orange box) occurred on the middle data qubit prior to expansion, as shown in Fig.~\ref{fig: error propagation}(a). After Stage 1, the error propagates onto two data qubits of the intermediate regular code shown in Fig.~\ref{fig: error propagation}(b). Following Stage 2, the original $X$ error spreads to four physical $X$ errors. However, this higher-weight error is equivalent to two $X$ errors (red boxes) in Fig.~\ref{fig: error propagation}(c). These errors anti-commute with two $Z$ stabilizers (red stars), leading to two detection events. This demonstrates that even after propagation, errors continue to trigger at most two syndrome changes, supporting the use of matching-based decoders.

\section{Total number of gates to prepare surface codes unitarily} \label{app: total gate count}
Assume we start from $\rot 3$ as the initial code, and unitarily prepare the final code of distance $D=2^m+1$ for some $m\in\mathbb N$. In the local encoder, it takes $8d+4$ CX gates to expand $\rot d$ to $\rot{d+2}$. The total number of gates is $N_\mathrm{loc}=\sum_{d=3,5,...}^{D-2}(8d+4)=2D^2-2D-12$. For our non-local encoder, it takes $(6d^2-10d+4)$ gates to expand $\rot{d}$ to $\rot{2d-1}$. The code distance in the $i$-th step of grown is denoted by $d_i=2^i+1$. The total number of gates is $N_\mathrm{nonloc}=\sum_{i=1}^{m-1}(6d_i^2-10d_i+4)=2\cdot 4^m+2^{m+1}-12=2D^2-2D-12$. Thus the total number of gates required are the same for both local or non-local encoder.

% \section{An optimized atom-movement protocol for our encoding scheme} \label{app: atom movements}
% \begin{itemize}
%     \item  From rot(3) to rot(5), hold all 4+12=16 ancillas on a moving trap.
%     \item Any CX, which requires H, can be done by global single qubit pulse and qubit detuning
%     \item rot to reg
%         \SubItem{Gate 1: move whole trap diagonally and shine 2Q pulse}
%         \SubItem{Gate 2: two parts for two different moving directions, need to detune some qubits to avoid unwanted 2Q gates.}
%     \item reg to rot
%         \SubItem{Gate 3: move whole lattice upward for gates between qubits in two traps; at the same time bring pairs of rows in the moving lattice closer together for gates between qubits within the moving trap.}
%         \SubItem{By symmetry, gate 4 can be done in a same way, rotated by 90 deg.}
%     \item Total: 5 2Q gate time. One load/unload
% \end{itemize}

\bibliography{references}

\end{document}